\documentclass[sort&compress,5p]{elsarticle}

\usepackage{bm}
\usepackage{amsmath}
\usepackage{amssymb}
\usepackage{graphicx}
\usepackage{subfigure}
\usepackage{float}
\usepackage[usenames,dvipsnames]{color}
\definecolor{darkblue}{RGB}{0,0,196}
\usepackage[colorlinks=true,linkcolor=darkblue,citecolor=darkblue,urlcolor=darkblue]{hyperref}

\usepackage{setspace}
\usepackage{footmisc}
\usepackage[makeroom]{cancel}

\newcommand{\intdP}{\int\!dP}

\def\be{\begin{equation}}
\def\ee{\end{equation}}
\def\ba{\begin{eqnarray}}
\def\ea{\end{eqnarray}}

\journal{Physics Letters B}

\begin{document}

\begin{frontmatter}



\title{Bulk observables at 5.02 TeV using quasiparticle anisotropic hydrodynamics}


\author{Mubarak Alqahtani} 
\address{Department of Basic Sciences, College of Education, Imam Abdulrahman Bin Faisal University, Dammam 34212, Saudi Arabia}

\author{Michael Strickland} 
\address{Department of Physics, Kent State University, Kent, OH 44242 United States}

\begin{abstract}
We present comparisons between 3+1D quasiparticle anisotropic hydrodynamics (aHydroQP) predictions for a large set of bulk observables and experimental data collected in 5.02 TeV Pb-Pb collisions.  We make aHydroQP predictions for identified hadron spectra, identified hadron average transverse momentum, charged particle multiplicity as a function of pseudorapidity, the kaon-to-pion ($K/\pi$) and proton-to-pion ($p/\pi$) ratios, and integrated elliptic flow.  We compare to data collected by the ALICE collaboration in 5.02 TeV Pb-Pb collisions.  We find that these bulk observables are quite well described by aHydroQP with an assumed initial central temperature of $T_0=630$ MeV at $\tau_0 = 0.25$ fm/c and a constant specific shear viscosity of $\eta/s=0.159$ and a peak specific bulk viscosity of $\zeta/s = 0.048$.  In particular, we find that the momentum dependence of the kaon-to-pion ($K/\pi$) and proton-to-pion ($p/\pi$) ratios reported recently by the ALICE collaboration are extremely well described by aHydroQP in the most central collisions. 
\end{abstract}







\end{frontmatter}

\section{Introduction}
\label{sec:intro}

At high-temperatures one expects hadronic matter to undergo a phase transition to a quark-gluon plasma (QGP) in which the appropriate degrees of freedom are quarks and gluons rather than hadrons.  The phase transition from hadronic matter to QGP is associated with both the restoration of chiral symmetry and deconfinement of the quarks and gluons.  Direct numerical calculations of the QGP phase transition temperature using lattice QCD have found that the transition is a smooth crossover with a crossover temperature of \mbox{$T_c \sim 155$ MeV}~\cite{Bazavov:2013txa,Borsanyi:2016bzg}.  To produce the QGP in the lab, experimentalists at the Relativistic Heavy Ion Collider (RHIC) and the Large Hadron Collider (LHC) collide ultrarelativistic nuclei in order to create a short-lived QGP with a lifetime on the order of 12 fm/c in central 5.02.TeV Pb-Pb collisions.  Analysis of the data produced in the last decades has shown that many aspects of the collective behavior observed in high-energy heavy-ion collisions are well-described by relativistic viscous hydrodynamics with an equation of state that takes into account the transition between hadronic and partonic degrees of freedom~\cite{Huovinen:2001cy,Romatschke:2007mq,Ryu:2015vwa,Niemi:2011ix,Averbeck:2015jja,Jeon:2016uym,Romatschke:2017ejr}.

In viscous hydrodynamics approaches one typically starts from the assumption that the non-equilibrium corrections to the dynamics, e.g. shear and bulk viscous tensors, are small relative to the equilibrium contributions to the energy-momentum tensor. One of the challenges such approaches face is that at early times,  $\tau < 1$ fm/c,  the QGP created in heavy-ion collisions can suffer from quite large deviations from thermal equilibrium.  These deviations can be large enough that the resulting one-particle distributions become negative in large regions of phase space \cite{1410.5786,1605.02101}.   Additionally, one finds that in such models the diagonal components of the energy-momentum tensor (anisotropic local rest frame pressures) can become negative even in cases where the underlying model being described by viscous hydrodynamics can never possess negative pressures, e.g. kinetic theory in relaxation time approximation.  These problems are particularly worrisome at early-times and near the transverse/longitudinal edges of the system \cite{Martinez:2009mf,Florkowski:2016kjj}.

The emergence of these problems is related to the fact the viscous hydrodynamics equations of motion are typically truncated at second-order in an expansion in terms of the inverse Reynolds number and Knudsen numbers of the system.  While it is formally possible to go beyond second order viscous hydrodynamics to third and higher orders in order to improve the treatments, an alternative idea was introduced in Refs.~\cite{Florkowski:2010cf,Martinez:2010sc} called anisotropic hydrodynamics (aHydro) in which no truncation in the inverse Reynolds number is performed.  This non-perturbative treatment of the response of the system to large gradients has been seen to result in better agreement between aHydro and exact solutions than seen with traditional second-order approaches~\cite{Florkowski:2013lza,Florkowski:2013lya,Florkowski:2014sfa,Denicol:2014tha,Denicol:2014xca,Nopoush:2014qba,Baym:1984np,Baym:1985tna,Heiselberg:1995sh,Wong:1996va,Strickland:2018ayk,Strickland:2019hff,Alalawi:2020zbx,Almaalol:2020rnu}.  

Since its inception, the original anisotropic hydrodynamics framework proposed in Refs.~\cite{Florkowski:2010cf,Martinez:2010sc} has been extended to  full 3+1-dimensional (3+1D) hydrodynamics including a realistic equation of state taken from lattice QCD calculations.  In addition, both shear and bulk viscouse correction plus an infinite set of implicit higher-order transport coefficients are included, as there is no truncation in inverse Reynolds number \cite{Ryblewski:2010ch,Florkowski:2011jg,Martinez:2012tu,Ryblewski:2012rr,Bazow:2013ifa,Tinti:2013vba,Nopoush:2014pfa,Tinti:2015xwa,Bazow:2015cha,Strickland:2015utc,Alqahtani:2015qja,Molnar:2016vvu,Molnar:2016gwq,Alqahtani:2016rth,Bluhm:2015raa,Bluhm:2015bzi,Alqahtani:2017jwl,Alqahtani:2017tnq} (for a recent review, see Ref.~\cite{Alqahtani:2017mhy}).  In Refs.~\cite{Alqahtani:2017jwl,Alqahtani:2017tnq,Almaalol:2018gjh,1807.05508,1811.01856,2007.04209} the resulting 3+1D quasiparticle anisotropic hydrodynamics code (aHydroQP) was used to make model to data comparisons for 2.76 TeV Pb-Pb collisions and 200 GeV Au-Au collisions.  These prior studies found quite good agreement between aHydroQP and many heavy-ion observables such as the identified hadron spectra,  mean transverse momentum of identified hadrons, multiplicities, the elliptic flow, and the HBT radii. For all of these observables, the aHydroQP model was able to describe the data quite reasonably over a broad range of centrality bins.

In this work, we continue our comparisons with heavy-ion experimental data, this time for 5.02 TeV collisions.  We present predictions for identified hadron spectra and their ratios, the charged particle multiplicity, identified hadron average transverse momentum, and integrated elliptic flow.  We compare our aHydroQP predictions with data collected by the ALICE experiment and find that the agreement with data is quite good.  In particular, we find that the momentum dependence of the kaon-to-pion ($K/\pi$) and proton-to-pion ($p/\pi$) ratios reported recently by the ALICE collaboration are extremely well described by aHydroQP in 0-5\% centrality collisions out to transverse momentum of 2.5 GeV.  The resulting initial temperature extracted for the QGP in 5.02 TeV collisions is $T_0 = 630$ MeV at $\tau_0 = 0.25$ fm/c and the specific shear viscosity found to give best agreement with the data is $\eta/s =  0.159$, with a corresponding peak specific bulk visocity of $\zeta/s = 0.048$.  The results presented here provide a standard point of reference for using aHydroQP in the calculation of other heavy-ion observables.  Even prior to this manuscript the results of this tuning have been used, e.g. to study the nuclear modification factor $R_{AA}$ and elliptic flow $v_2$ of bottomonium states in 5.02 TeV Pb-Pb collisions \cite{2007.03939,2007.10211}.

The structure of our paper is as follows.  In Sec.~\ref{sec:model}, we review the basics of the 3+1D quasiparticle anisotropic hydrodynamics model.  In Sec.~\ref{sec:results},  we present the predictions of the aHydroQP model for 5.02 TeV collisions and compare them to experimental data for many heavy-ion observables. Sec.~\ref{sec:conclusions} contains our conclusions and an outlook for the future. 

\section{Model}
\label{sec:model}

\begin{figure*}[t!]
\centerline{
\includegraphics[width=0.83\linewidth]{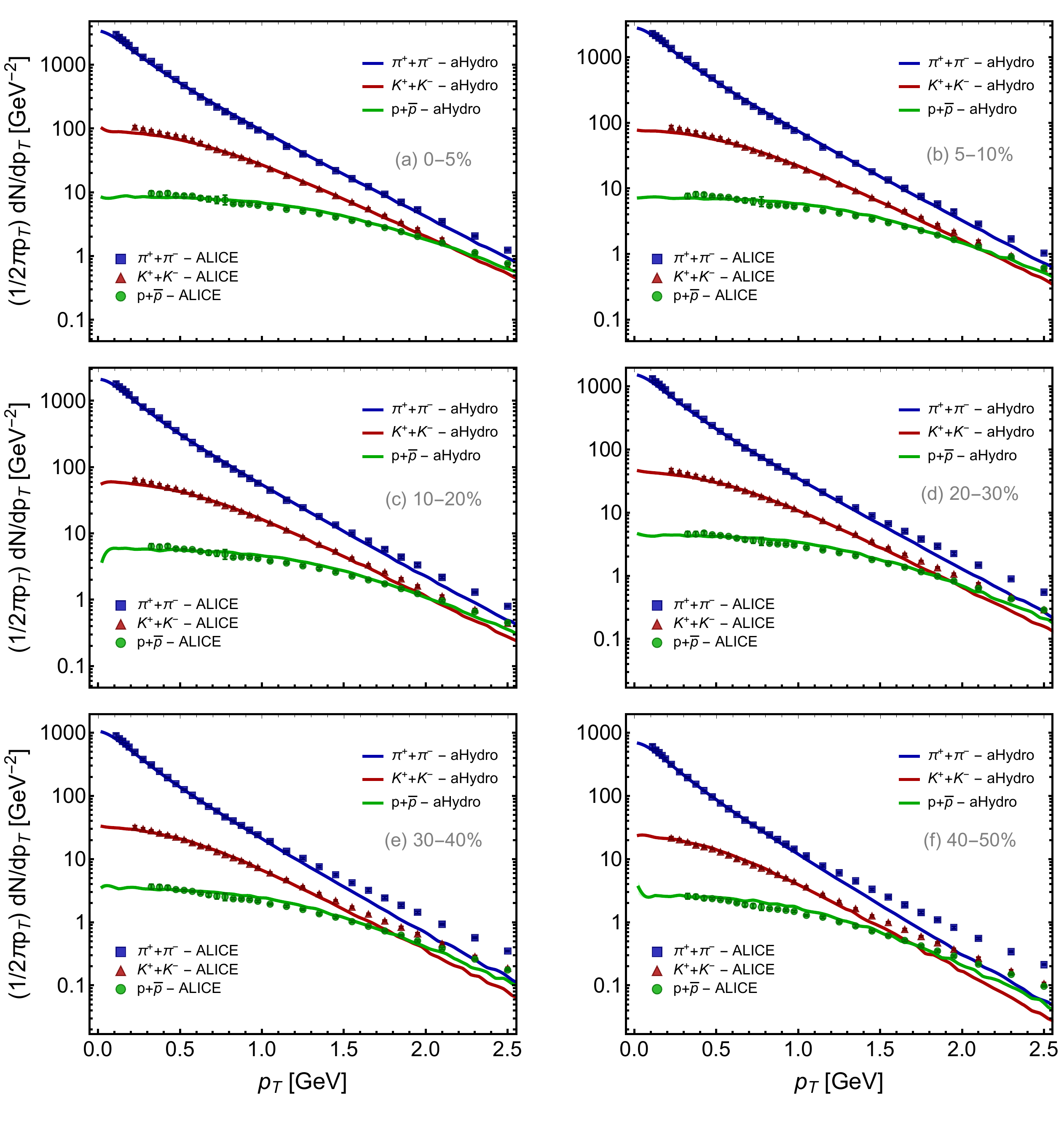}}
\caption{Combined transverse momentum spectra of pions, kaons and protons for 5.02 TeV Pb-Pb collisions in different centrality classes. The solid lines are the predictions of 3+1D aHydroQP and the points are experimental results from the ALICE Collaboration~\cite{1910.07678}. }
\label{fig:spectra}
\end{figure*}

The evolution of the medium is performed using the first and second moments of Boltzmann equation for a system with temperature-dependent quasiparticle masses~\cite{Alqahtani:2015qja}
\be
p^\mu \partial_\mu f(x,p)+\frac{1}{2}\partial_i m^2\partial^i_{(p)} f(x,p)=-C[f(x,p)] \, .
\label{eq:boltzmanneq}
\ee
The first and second moments of Eq.~\eqref{eq:boltzmanneq} give
\be
\partial_\mu  T^{\mu \nu} =0\, \label{eq:1stmoment} ,
\ee
\be
\partial_\alpha  I^{\alpha\nu\lambda}- J^{(\nu} \partial^{\lambda)} m^2 =-\intdP \, p^\nu p^\lambda{\cal C}[f]\, \label{eq:2ndmoment} ,
\ee
where
\be
J^\mu = \int dP \, p^\mu f \, ,
\ee
\be
T^{\mu\nu} = \int dP \, p^\mu p^\nu  f \, ,
\ee
and 
\be
I^{\mu\nu\lambda} = \int dP \, p^\mu p^\nu p^\lambda f \, ,
\ee
with $\int dP = N_\text{dof} \int \frac{d^3{\bf p}}{(2\pi)^3} \frac{1}{E}$ being the Lorentz invariant integration measure.

In the aHydroQP approach, the mass is a function of temperature which can be obtained from lattice QCD calculations of the entropy density in order to enforce a realistic equation of state~\cite{Alqahtani:2015qja}.  The collisional kernel  $C[f(x,p)]$ in aHydroQP is  taken in the relaxation-time approximation (RTA). 
\be
C[f] = - \frac{p \cdot u}{\tau_{\rm eq}(T)} [ f - f_\text{eq}(T) ] \, ,
\ee
where $u^\mu$ is the four-velocity of the fluid local rest frame and $\tau_{\rm eq}(T)$ is the temperature- (and hence time-) dependent relaxation time \cite{Alqahtani:2015qja}.

We obtain the necessary dynamical equations by taking projections of the Boltzmann equation.  For this purpose, one needs to first specify the form of the underlying one-particle distribution function. In aHydroQP, the distribution function is taken to be anisotropic in momentum space, with only diagonal momemtum-space anisotropy parameters, and having the form
\be
f(x,p) =  f_{\rm eq}\!\left(\frac{1}{\lambda}\sqrt{\sum_i \frac{p_i^2}{\alpha_i^2} + m^2}\right) .
\label{eq:fform}
\ee
This distribution function reduces back to an equilibrium distribution function with temperature $T$ when $\alpha_i=1$ and  $\lambda$ = T.   For details of the derivation of the dynamical equations for aHydroQP we refer readers to Refs.~\cite{Nopoush:2014pfa,Alqahtani:2015qja,Alqahtani:2016rth,Alqahtani:2017mhy}.

\begin{figure*}[t!]
\centerline{
\includegraphics[width=0.75\linewidth]{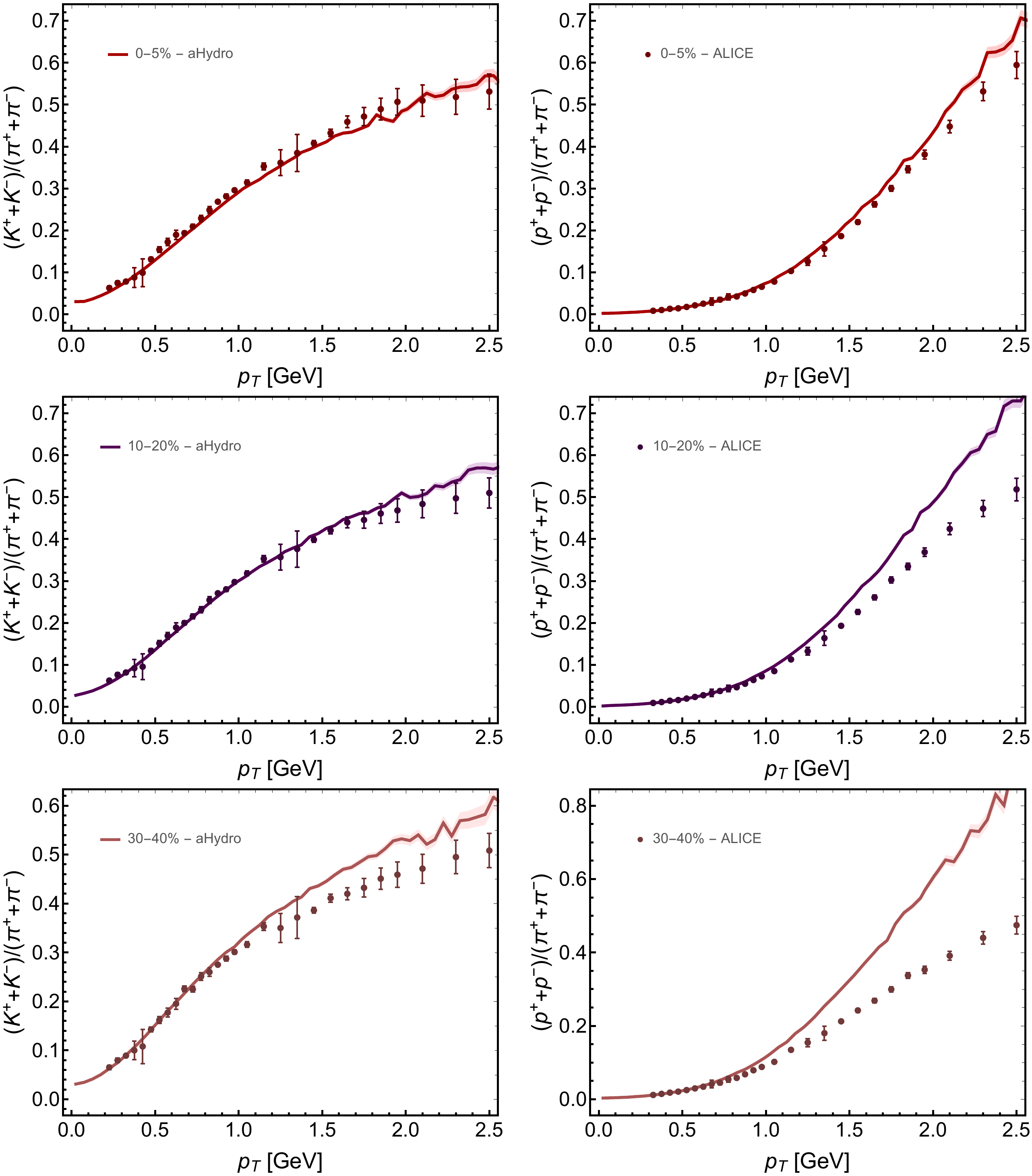}}
\caption{ The $K/\pi$ (left) and $p/\pi$ (right) ratios as a function of $p_T$ measured in Pb-Pb collisions at 5.02 TeV in different centrality classes. Solid lines are predictions of aHydroQP model where symbols with error bars are experimental data from Ref.~\cite{1910.07678}.}
\label{fig:ratios}
\end{figure*}

Using the aHydroQP dynamical equations we allow the system to evolve until reaching the freeze-out temperature $T_{\rm FO}=130$ MeV where a hypersurface is constructed at a constant energy-density.  On this hypersurface we convert the underlying hydrodynamic evolution results for the flow velocity, the anisotropy parameters, and the scale $\lambda$ into explicit `primioridial' hadronic distribution functions using a generalized Cooper-Frye prescription \cite{Alqahtani:2017mhy}.   For this purpose, we use a customized version of THERMINATOR 2 ~\cite{1102.0273} to perform the production and necessary decay(s) of the primordial hadrons.  Both the aHydroQP and modified THERMINATOR 2a codes are publicly available \cite{MikeCodeDB}.  Note that the freeze-out temperature used herein and all other parameters besides $T_0$ and $\eta/s$ were assumed to be the same as in our prior 2.76 TeV study \cite{Alqahtani:2017jwl,Alqahtani:2017tnq}.  Similarly, in order to have a meaningful comparison to results obtained previously at 200 GeV and 2.76 TeV, we use smooth Glauber type initial conditions with the central energy density scaling with the nuclear overlap profile.

\section{Results}
\label{sec:results}

In this section, we present comparisons of aHydroQP predictions with 5.02 TeV Pb-Pb collision data collected by the ALICE collaboration. We consider only two free parameters, the initial central temperature $T_0$ and the specific shear viscosity $\eta/s$.\footnote{Note that aHydroQP also includes bulk viscous effects, however, within the relaxation time approximation, the bulk viscosity as a function of temperature is fixed once one specifies the shear viscosity~\cite{Alqahtani:2015qja,Alqahtani:2017jwl,Alqahtani:2017tnq}.} We fix these two parameters by fitting the spectra of pions, kaons, and protons in both the 0-5\% and 30-40\% centrality classes. The parameters obtained from the spectra fit were: $T_0=630$ MeV and $\eta/s=0.159$.  The initial temperature obtained is higher than found at 2.76 TeV, $T_0^\text{2.76 TeV} = 600 $ MeV, by 5\% \cite{Alqahtani:2017tnq}.  The best fit value for $\eta/s$ is the same as was found at 2.76 TeV \cite{Alqahtani:2017tnq}.

\begin{figure*}[t!]
\centerline{
\includegraphics[width=0.75\linewidth]{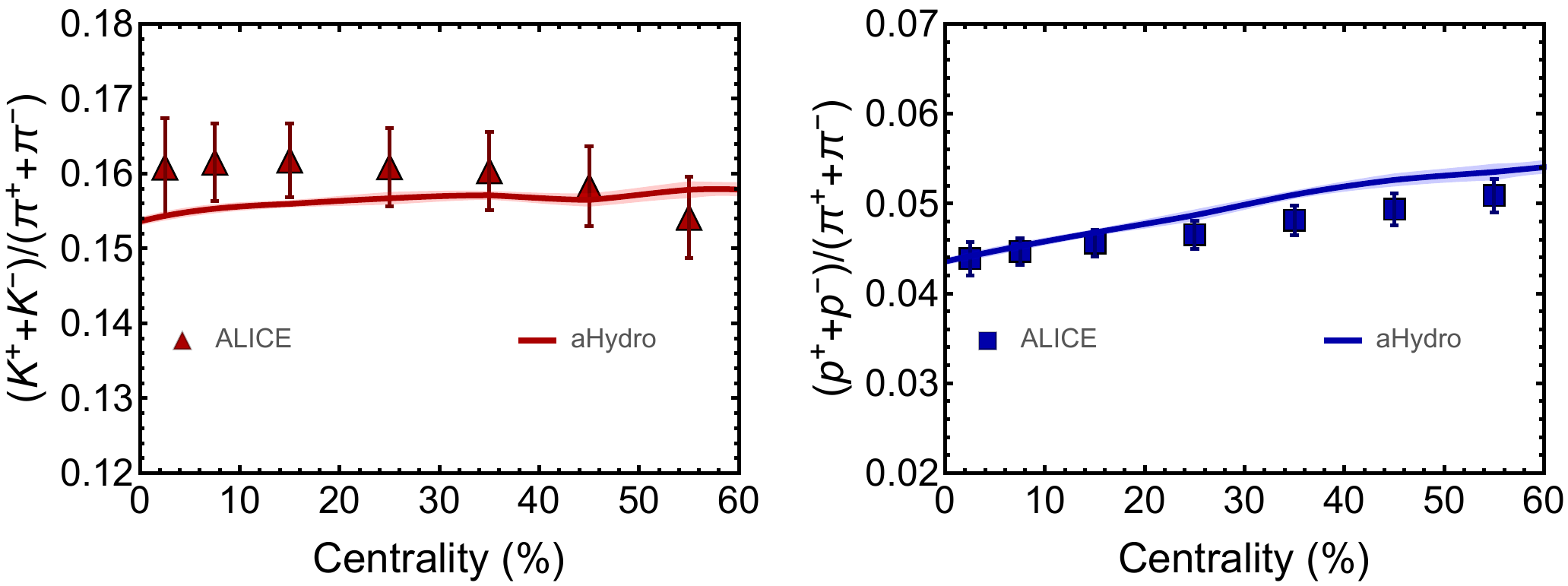}}
\caption{Transverse-momentum integrated $K/\pi$ (left) and $p/\pi$ (right) ratios as a function of centrality measured in Pb-Pb collisions at 5.02 TeV. Solid lines are predictions of aHydroQP model where symbols with error bars are experimental data from Ref.~\cite{1910.07678}.   }
\label{fig:ratios-cent}
\end{figure*}

\begin{figure*}[t!]
\centerline{
\includegraphics[width=0.8\linewidth]{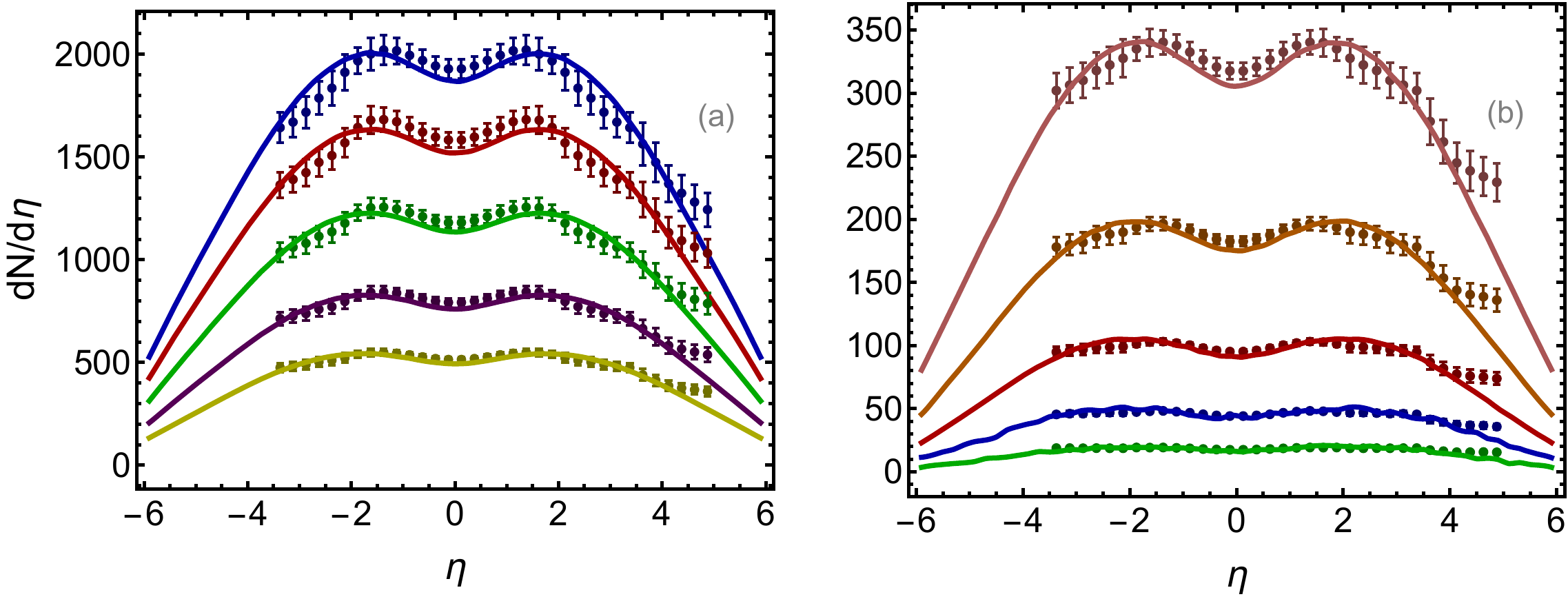}}
\caption{ The charged-particle pseudorapidity density in Pb-Pb collisions at 5.02 TeV obtained by aHydroQP model (solid lines) as a function of pseudorapidity $\eta$. Different centrality classes are shown in both panels (a) and (b) starting from more central 0-5\% (blue) to more peripheral 80-90\% (green). Data are from ALICE Collaboration Ref.~\cite{1612.08966}. }
\label{fig:dNdeta}
\end{figure*}

We begin by presenting aHydroQP predictions for the transverse momentum distribution of identified hadrons in 5.02 TeV Pb-Pb collisions.  We will compare our aHydroQP predictions with experimental data from the ALICE collaboration~\cite{1704.06030,1910.07678}. In Fig.~\ref{fig:spectra}, we show the combined spectra of pions, kaons, and protons as a function of transverse momentum in six different centrality classes. In more central collisions, aHydroQP shows very good agreement with the data as shown in Fig.~\ref{fig:spectra}a. On the other hand, for more peripheral collisions the agreement is good only for $p_T  \lesssim 1$ GeV as can be seen in, e.g., Fig.~\ref{fig:spectra}f. 

In Fig.~\ref{fig:ratios}, the $K/\pi$ (left column) and $p/\pi$ (right column) ratios are shown as a function of $p_T$ in three different centrality classes and once again compared to experimental data. The agreement between aHydroQP and the data at $p_T \lesssim 1$ GeV is very good in all centrality bins. In the 0-5\% central class, for the $K/\pi$ ratio, the agreement between aHydroQP and the data extends up to $p_T  \sim 2.5$ GeV, while for $p/\pi$ is extends up to $p_T  \sim 1.5$ GeV. The aHydroQP predictions for the $K/\pi$ and $p/\pi$ rations as a function of centrality are shown in Fig.~\ref{fig:ratios-cent}, in the left and right panels, respectively. In both panels, we see that aHydroQP is able to describe the ratios reported by the ALICE collaboration well over a broad range of centralities.

In Fig.~\ref{fig:dNdeta}, we present the aHydroQP prediction for the charged-particle pseudorapidity density in Pb-Pb collisions at 5.02 TeV along with data provided by the ALICE collaboration~\cite{1612.08966} .   In all centrality bins shown, aHydroQP describes the data quite well over a broad range of pseudorapidity.   At high rapidities we notice some differences from the data where there are indicates of a more slow decrease.  This was not the case at 2.76 TeV where, in the most central class, the agreement between aHydroQP and experimental data extended out to $|\eta| \sim 5$.  Overall, however, we see good agreement at central rapidities in all centrality classes considered in Fig.~\ref{fig:dNdeta}.

\begin{figure*}[t!]
\centerline{
\includegraphics[width=0.375\linewidth]{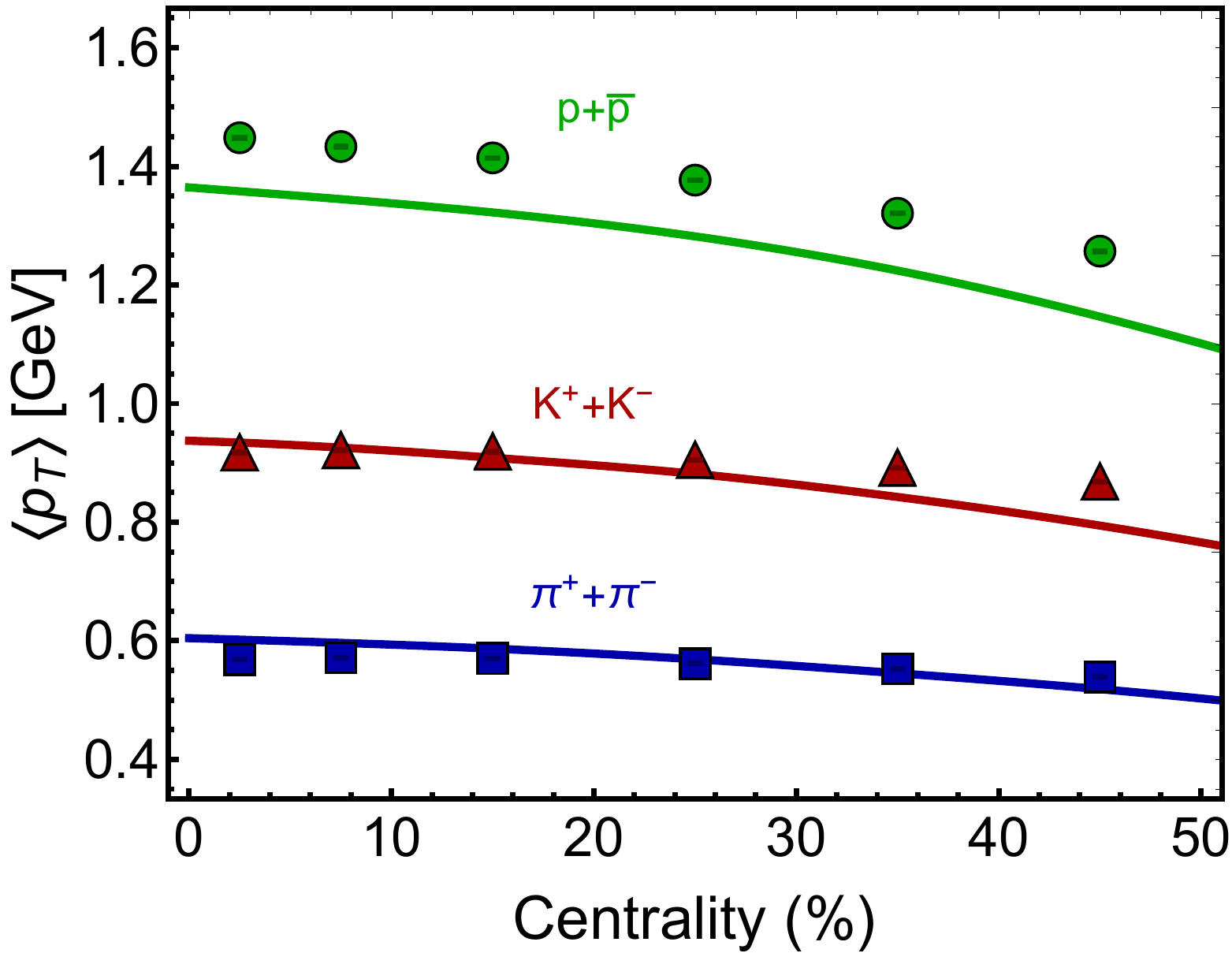}\hspace{1cm}
\includegraphics[width=0.4\linewidth]{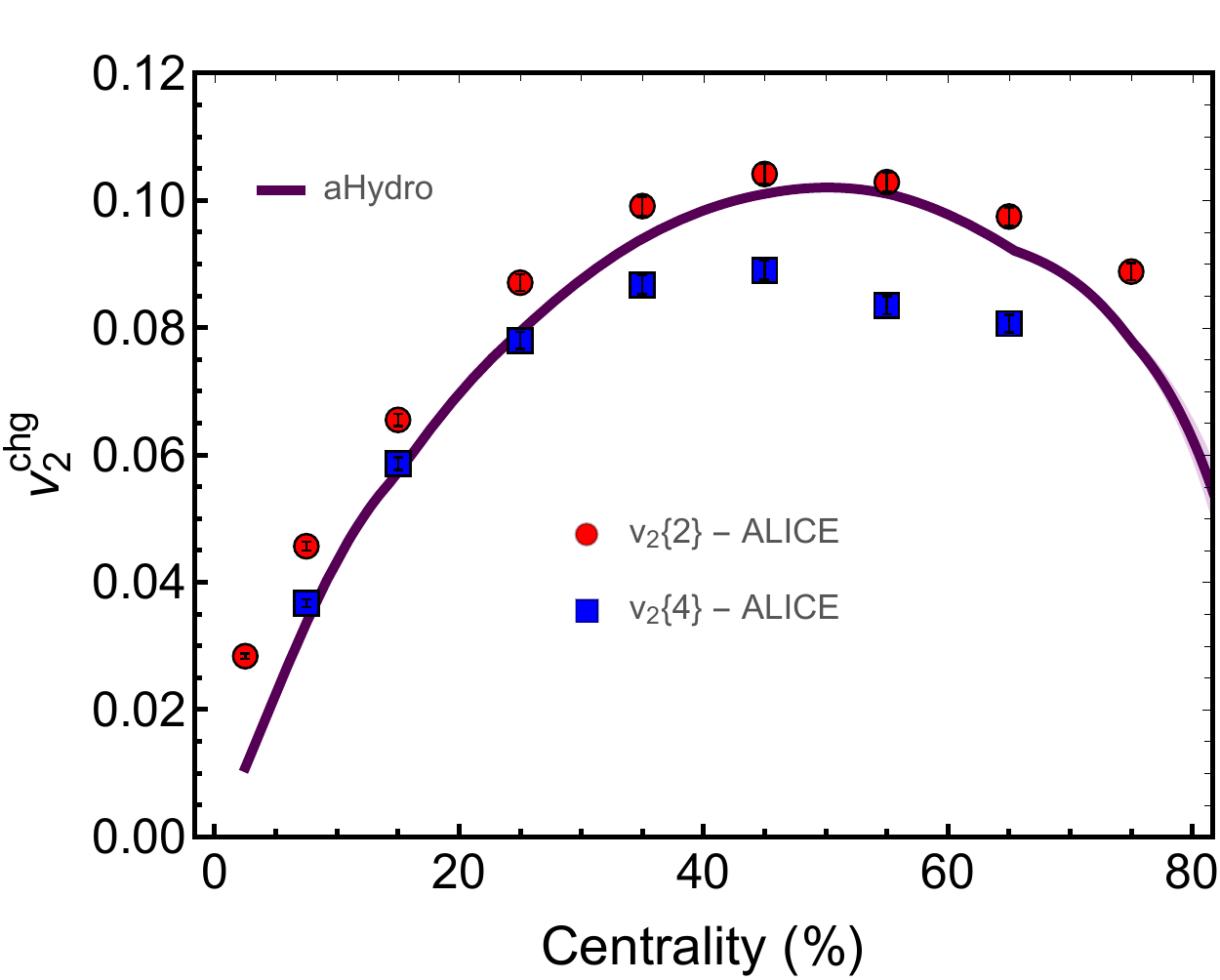}}
\caption{ In panel (a), identified particle mean transverse momentum vs. centrality is shown in 5.02 TeV Pb+Pb collisions where data are from ALICE Collaboration Ref.~\cite{1910.07678}, while in panel (b), the centrality dependence of the elliptic flow $v_2$ of charged particles in 5.02 TeV Pb-Pb collisions is shown where data are from Ref.~\cite{1602.01119}. }
\label{fig:ptv2}
\end{figure*}

Next we turn to the left panel of Fig.~\ref{fig:ptv2} in which present aHydroQP predictions for the mean transverse momentum $\langle p_T \rangle$ of pions, kaons and protons.  The aHydroQP results are compared to experimental data from the ALICE collaboration \cite{1910.07678}.  For both pions and kaons, we see good agreement between aHydroQP and the experimental data at all centralities, however, aHydroQP seems to underestimate the mean $p_T$ for protons.  We have no immediate explanation for why there is such a discrepancy, but we do note that a similar discrepancy exists in state-of-the-art second-order viscous hydrodynamics calculations~\cite{2005.14682}.

Finally, in Fig.~\ref{fig:ptv2} (right panel), we present the aHydroQP predictions for the integrated flow as a function of centrality.  Since we used smooth initial conditions, the event plane is known and we computed $v_2$ using the $\langle \cos(2\phi) \rangle$ for all hadrons.  The aHydroQP predictions are compared to ALICE data for $v_2\{2\}$ and $v_2\{4\}$ reported in Ref.~\cite{1602.01119}.  As can be seen from this comparison, the aHydroQP predictions agree well with the experimentally measured $v_2\{4\}$ in the most central bins ($< 25 \%$) and are closer to $v_2\{2\}$ at higher centralities.  

\section{Conclusions and outlook}
\label{sec:conclusions}

In this work, we continued our comparisons of aHydroQP with experimental data. In the past, we presented comparisons with data at 2.76 TeV Pb-Pb collisions \cite{Alqahtani:2017jwl,Alqahtani:2017tnq} and 200 GeV Au-Au collisions \cite{Almaalol:2018gjh,2007.04209}.  Herein we made theory to data comparisons between aHydroQP and data collected by the ALICE collaboration using 5.02 TeV Pb-Pb collisions.  We presented aHydroQP predictions for a large set of bulk observables and found quite reasonable agreement with experimental data using a central temperature of $T_0 = 630$ MeV at $\tau = 0.25$ fm/c and a specific shear viscosity of $\eta/s = 0.159$.  Comparing to our previous studies at 2.76 TeV \cite{Alqahtani:2017jwl,Alqahtani:2017tnq}, all other parameters besides $T_0$ and $\eta/s$ were assumed to be the same.

We made aHydroQP predictions for identified hadron spectra, identified hadron average transverse momentum, charged particle multiplicity as a function of rapidity, the kaon-to-pion ($K/\pi$) and proton-to-pion ($p/\pi$) ratios, and integrated elliptic flow.  In all of these comparisons, aHydroQP was quite successful in describing the experimental data in 5.02 TeV Pb-Pb collisions. In particular, we find that the momentum dependence of the kaon-to-pion ($K/\pi$) and proton-to-pion ($p/\pi$) ratios reported recently by the ALICE collaboration are extremely well described by aHydroQP in the most central collisions.  In a followup paper, due to the increased statistics required, we intend to present aHydroQP predictions for Hanbury Brown-Twiss radii and compare aHydroQP predictions for the identified hadron elliptic flow for pions, kaons, and protons with experimental data~\cite{1805.04390}. 

We note, in closing, that the the initial state model (Glauber model) and assumed collision kernel (RTA) used herein are rather simple.  As a result of the smooth initial condition assumed, we do not correctly reproduce the $v_2^{\rm chg}$ in the most central collisions.  The aHydroQP code allows for fluctuating initial conditions and we plan to report on the results of such simulations in a forthcoming paper.   One of the challenges with using, e.g. IPGlasma-type, fluctuating initial conditions is that these types of initial conditions can possess large numbers of cells in which there are negative total pressures in the local rest frame, which is incompatible with the kinetic-theory based assumptions underpinning aHydroQP.  With respect to the collision kernel, a framework for including realistic collisional kernels in the aHydro framework was introduced in Refs.~\cite{Almaalol:2018ynx,Almaalol:2018jmz}.  It will be interesting to see if aHydroQP results are sensitive to the choice of the collisional kernel.  

Finally, we mention that another assumption made herein was that the anisotropy tensor is diagonal in the local-rest frame.  Although this is justified by the smallness of the off-diagonal contributions, it is desirable to have a complete treatment which includes the off-diagonal contributions in a non-perturbative manner.  Such a scheme was introduced in Ref.~\cite{Nopoush:2019vqc} and is currently being implemented.

\section*{Acknowledgments}

M. Alqahtani is supported by the Deanship of Scientific Research at the Imam Abdulrahman Bin Faisal University under grant number 2020-080-CED. M. Strickland was supported by the U.S. Department of Energy, Office of Science, Office of Nuclear Physics under Award No. DE-SC0013470. This research in part utilized Imam Abdulrahman Bin Faisal (IAU)'s Bridge HPC facility, supported by IAU Scientific and High Performance Computing Center~\cite{Bridge}.

\bibliographystyle{elsarticle-num}
\bibliography{fivetev}

\end{document}